\documentclass[%
reprint,
showpacs,reprintnumbers,
 amsmath,amssymb,
 aps,
prl,
superscriptaddress,
showpacs, showkeys]{revtex4-1}
\usepackage{amsmath}
 \usepackage{subfigure}
\usepackage{textgreek}
\usepackage{subfloat}
\usepackage{color}
\usepackage{breqn}
\usepackage{graphicx}
\usepackage{dcolumn}
\usepackage{bm}
\usepackage{hyperref}
\usepackage{textcomp}
\usepackage{xr}
\hypersetup{bookmarksnumbered, pdfpagemode=UseOutlines, 
colorlinks=true, citecolor=blue, filecolor=blue, linkcolor=blue, urlcolor=blue}
 \newcommand{\beginsupplement}{%
         \setcounter{table}{0}
         \renewcommand{\thetable}{S\arabic{table}}%
         \setcounter{figure}{0}
         \renewcommand{\thefigure}{S\arabic{figure}}
         \setcounter{equation}{0}
         \renewcommand{\theequation}{S\arabic{equation}}%
      }

 \makeatletter
 \let\cat@comma@active\@empty
 \makeatother

\graphicspath{/home/sid/Dropbox/6th_paper_WS2_susbstrate/paper_fig
}

\begin{document}

\preprint{}

\title{Nonlinear analog spintronics with van der Waals heterostructures}

\author{S. Omar} 
\thanks{corresponding author}
\email{s.omar@rug.nl}
\affiliation{The Zernike Institute for Advanced Materials, University of Groningen, Nijenborgh 4, 9747 AG, Groningen, The Netherlands}
\author{M. Gurram}
\affiliation{The Zernike Institute for Advanced Materials, University of Groningen, Nijenborgh 4, 9747 AG, Groningen, The Netherlands}%
\author{K. Watanabe}
\affiliation{National Institute for Material Science, 1-1 Namiki, Tsukuba, 305-044, Japan}%
\author{T. Taniguchi}
\affiliation{National Institute for Material Science, 1-1 Namiki, Tsukuba, 305-044, Japan}%
\author{M.H.D. Guimar\~{a}es}
\affiliation{The Zernike Institute for Advanced Materials, University of Groningen, Nijenborgh 4, 9747 AG, Groningen, The Netherlands}%
\author{B.J. van Wees}
\affiliation{The Zernike Institute for Advanced Materials, University of Groningen, Nijenborgh 4, 9747 AG, Groningen, The Netherlands}%
\date{\today}

\begin{abstract}
\end{abstract}

\maketitle
\textbf{The current generation of spintronic devices, which use electron-spin relies on linear operations for spin-injection, transport and detection processes. The existence of nonlinearity in a spintronic device is indispensable for spin-based complex signal processing operations.  Here we for the first time demonstrate the presence of electron-spin dependent nonlinearity in a spintronic device, and measure up to $4^{\text{th}}$ harmonic spin-signals via nonlocal spin-valve and Hanle spin-precession measurements. We demonstrate its application for analog signal processing  over pure spin-signals such as amplitude modulation and heterodyne detection operations which require nonlinearity as an essential element. Furthermore, we show that the presence of nonlinearity in the spin-signal has an amplifying effect on the  energy-dependent conductivity induced nonlinear spin-to-charge conversion effect. The interaction of the two spin-dependent nonlinear effects  in the spin transport channel  leads to a highly efficient detection of the spin-signal without using ferromagnets. These effects are measured both at 4K and room temperature, and are suitable for their applications as nonlinear circuit elements in the fields of advanced-spintronics and spin-based neuromorphic computing.
}


Nonlinear elements, such as transistors and diodes, led Shockley and coworkers \cite{shockley_$pensuremath-n$_1951} to lay the foundation for electronics revolution, and underlie the modern-day electronics. However, such elements lack in the field of spintronics. Major ideas in the field of spintronics thus far have suggested the possibility of achieving a gate operation employing, for example, a Datta-Das transistor \cite{datta_electronic_1990, gmitra_proximity_2017}. The possibility of spin-signal amplification and processing has not been explored experimentally and forms a more fundamental building block to replace conventional electronics with the spin-based analogues \cite{zeng_graphene-based_2011,acremann_amplifier_2008,wang_graphene_2014,flatte_theory_2003, fabian_spin-polarized_2004}.

The current generation of state-of-the-art spintronic devices can only execute linear operations. In such devices, the output differential spin-signal $v_{\text{s}}$ \cite{avsar_colloquium:_2019, gurram_electrical_2018} scales with the applied input ac charge current $i$, i.e.
\begin{equation}
v_{\text{s}}=p_{\text{inj}}R_{\text{s}}p_{\text{det}}i
\label{vs}
\end{equation} 
 Here, $p_{\text{inj(det)}}$, the differential spin injection(detection) efficiency and $R_{\text{s}}$, the effective spin resistance of the spin transport channel \footnote{$R_{\text{s}}=R_{\text{sq}}\lambda_{\text{s}}\exp(-\frac{L}{\lambda_{\text{s}}})/2w$ is the effective spin-resistance with channel sheet resistance $R_{\text{sq}}$ and spin relaxation length $\lambda_{\text{s}}$. $L$ and $w$ are the length (injector-detector separation) and width of the transport channel} are constant, and  thus the relation $v_{\text{s}}\propto i$ is established.   Therefore, a nontrivial operation requiring nonlinearity can not be executed.  
 
  Interestingly, the differential spin-injection efficiency $p_{\text{inj}}$ of ferromagnetic (FM) tunnel contacts with atomically flat and pinhole-free thin hBN flakes as a tunnel barrier \cite{gurram_spin_2016,lee_electron_2011} depends on the input dc bias current $I$  \cite{gurram_bias_2017, leutenantsmeyer_efficient_2018, zhu_probing_2018}, and renders them as a viable platform to demonstrate spin-dependent nonlinear effects.  

\subsection{Nonlinear spin-injection}

\begin{figure*}
\includegraphics[scale=0.9]{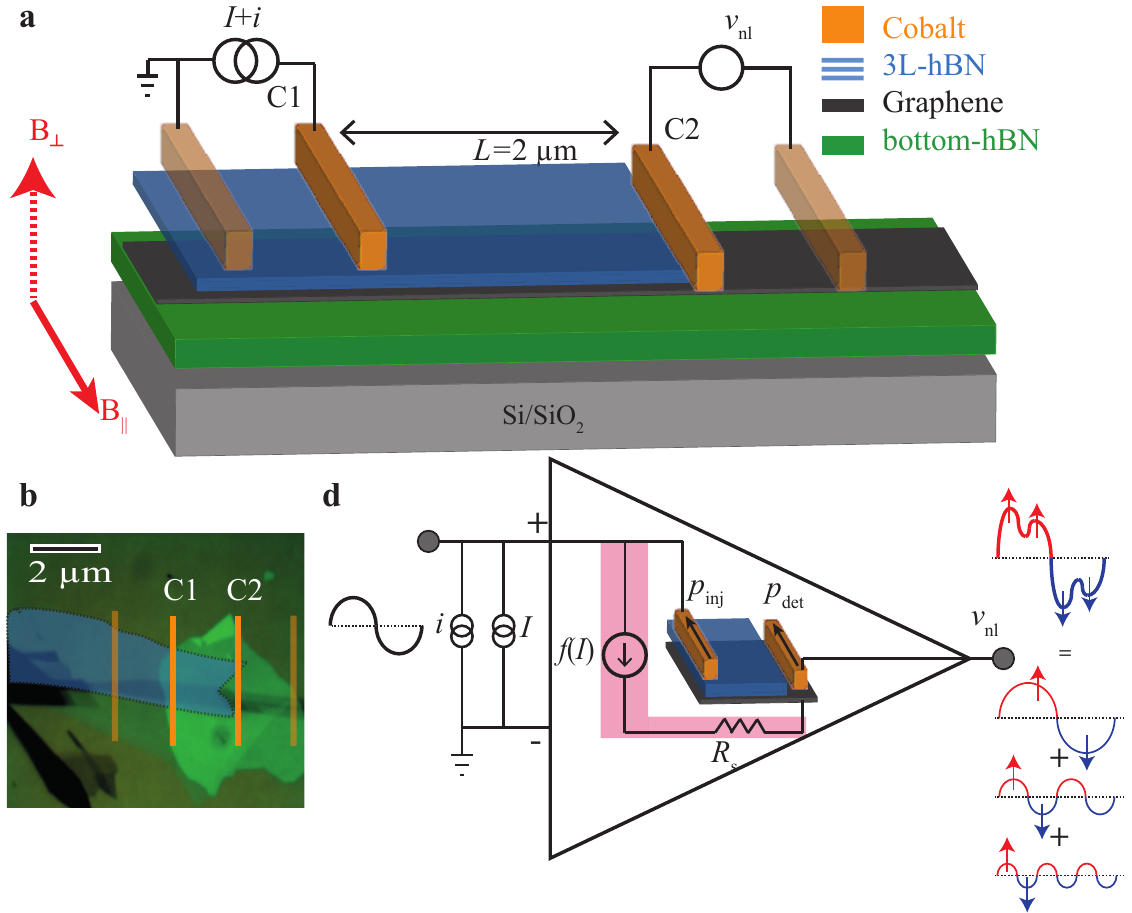}
\includegraphics[scale=1]{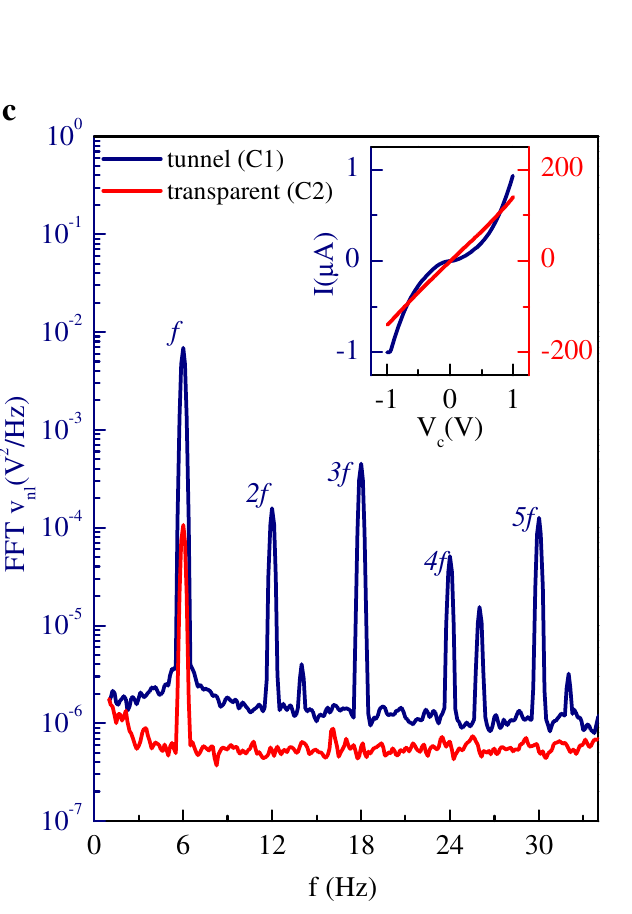}
\caption{\label{dev geo}\textbf{Device geometry and tunnel characteristics.} \textbf{a.} Graphene encapsulated between a thick hBN at the bottom and a 3L-hBN tunnel barrier on the top. The cobalt (inner) injector electrode C1 is located on top of the tunnel barrier and the (inner) detector electrode C2  is directly in contact with the graphene flake. The outer injector and detector electrodes (transparent orange) are far enough to be spin-sensitive, and serve as reference electrodes. A charge current $i+I$ is applied across C1 for spin-injection and a nonlocal ac voltage $v_{\text{nl}}$ is measured at C2 via the lock-in detection method. \textbf{b.} An optical image of the stack with the actual  positions of electrodes drawn schematically. The hBN-tunnel barrier is highlighted with false blue colour. \textbf{c.} An ac charge current $i$=50nA(20$\mu$A) at $f$=6 Hz is applied at C1 (C2) and the Fourier transform of the nonlocal signal measured at C2(C1) is plotted in blue(red).  In the inset, I-V characteristics of the tunnel contact C1 (blue) and the transparent contact C2 (red). \textbf{d.}  The concept of nonlinearity is presented schematically via a circuit diagram. A sinusoidal charge current $i$ along with a dc current $I$ is applied at the input of a nonlinear element (inside the triangle) and a distorted non-sinusoidal spin-signal is measured at the output. The harmonic components which construct the output signal are also shown. The equivalent circuit representing the bias dependent spin-injection ($p_{\text{inj}}=f(I)$) and spin transport ($R_{\text{s}}$) is highlighted in pink. \
}
\end{figure*}

We perform nonlinear spin-transport experiments on a van der Waals heterostructure of Graphene (Gr), encapsulated  between a thick boron nitride (hBN) substrate and a trilayer hBN tunnel barrier with ferromagnetic cobalt contacts as shown in Figs.~\ref{dev geo}(a,b). We start by characterizing the tunnelling behaviour of the contacts. Contacts with(out) the tunnel-barrier show nonlinear(linear) current-voltage characteristics [inset of Fig.~\ref{dev geo}(c)] for an applied dc charge current $I$ and the measured voltage $V_{\text{c}}$ across the contact in a three-probe measurement geometry.   Next, we probe the presence of nonlinear behaviour in the nonlocal signal $v_{\text{nl}}$ in a four-probe measurement geometry. For an input ac current $i$ at  frequency $f$=6 Hz, $v_{\text{nl}}$ is measured  using the scheme in Fig.~\ref{dev geo}(a), and its Fourier transform is shown in Fig.~\ref{dev geo}(c).  For a linear device, an applied current at a certain frequency should yield a voltage at the same frequency alone. The appearance of voltage at integral multiples of the input-current frequency, the so called higher harmonics, is a smoking gun signature of nonlinearity. In our measurements, higher harmonics at $2f, 3f, ...$  appear in $v_{\text{nl}}$ only when the tunnel contact C1 is used as an injector (blue spectrum in Fig.~\ref{dev geo}(c)), and thus underline the crucial role of tunnel contacts for introducing nonlinearity in $v_{\text{nl}}$.

\begin{figure*}
\includegraphics[scale=1]{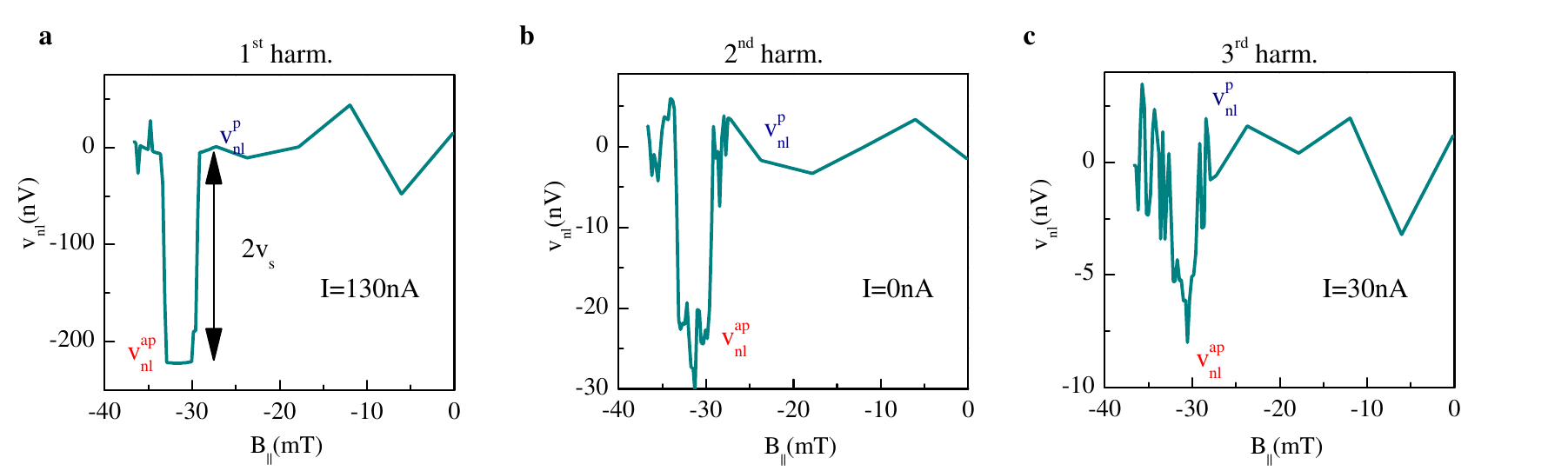}
\includegraphics[scale=1]{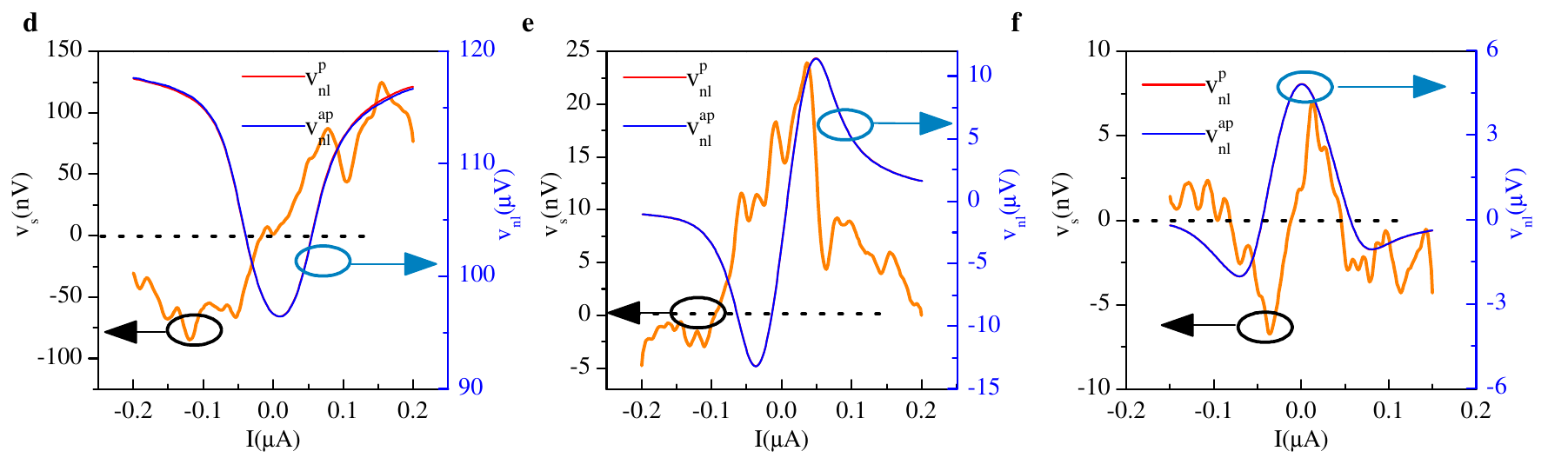}
\includegraphics[scale=1]{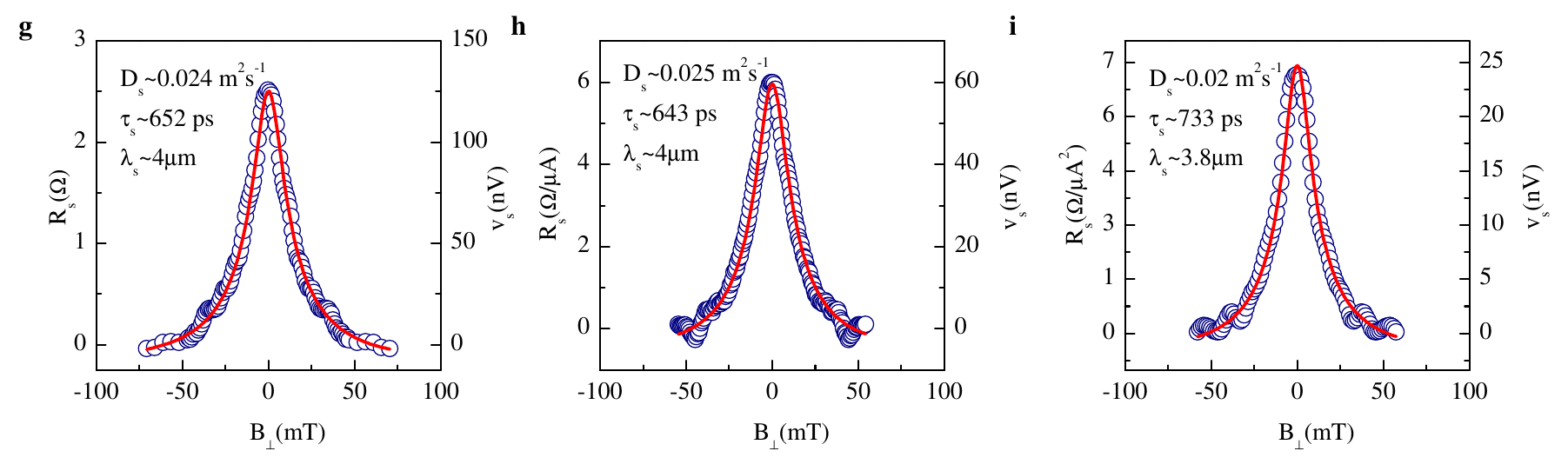}
\caption{\label{bias dep} \textbf{Higher harmonic spin-signals.} \textbf{a.}1$^{\text{st}}$, \textbf{b.} 2$^{\text{nd}}$ and \textbf{c.} 3$^{\text{rd}}$ harmonic spin-valve measurements.  \textbf{d-f.} spin-signal $v_{\text{s}}$ (orange) as a function of $I$ applied at C1, using the  measurement geometry in Fig.~\ref{dev geo}(a). The ac injection current $i$ is kept fixed at 50nA. $v_{\text{nl}}^{\text{p}}$ (red) and $v_{\text{nl}}^{\text{ap}}$ (blue) are the nonlocal signal $v_{\text{nl}}$ measured at the parallel and anti-parallel magnetization configurations of the injector-detector electrodes, respectively. \textbf{g.}1$^{\text{st}}$, \textbf{h.} 2$^{\text{nd}}$ and \textbf{i.} 3$^{\text{rd}}$ harmonic Hanle spin-signal $v_{\text{s}}$ as a function of out-of-plane magnetic field $B_{\perp}$. Hanle data is symmetrized in order to remove the linear background and is offset to zero. SV measurements in (a)-(f) are performed at RT, and Hanle curves in (g)-(h) are measured at 4K.}
\end{figure*}

The concept of nonlinear spintronic measurements is schematically demonstrated in Fig.~\ref{dev geo}(d). For an input charge current $i+I$ at a ferromagnetic contact, higher harmonics in the spin-signal $v_{\text{s}}$ are measured at the output, due to the nonlinearity in the spin-injection process, present in a spintronic device. To probe the spin-dependent origin of nonlinearity in $v_{\text{nl}}$, we perform bias dependent nonlocal spin-valve (SV) measurements \cite{gurram_bias_2017}. Here, we apply an ac+dc charge current $i+I$ and measure the $1^{\text{st}}$ harmonic response of $v_{\text{nl}}$ via the lock-in detection method. An in-plane magnetic field $B_{||}$ is swept to switch the magnetization-orientation of  C1 and C2 from parallel to anti-parallel and vice-versa using the connection scheme in Fig.~\ref{dev geo}(a). Via SV measurements, we obtain background free pure spin-signal $v_{\text{s}}=\frac{v_{\text{nl}}^{\text{p}}-v_{\text{nl}}^{\text{ap}}}{2}$, where $v_{\text{nl}}^{\text{ap}}(v_{\text{nl}}^{\text{p}})$ is the nonlocal signal $v_{\text{nl}}$ measured at the (anti-)parallel magnetization-direction alignment of the electrodes C1 and C2, as labeled in Fig.~\ref{bias dep}(a). In order to obtain the bias dependence of the spin-signal, we measure $v_{\text{nl}}^{\text{p(ap)}}$ as a function of $I$, as shown in Fig.~\ref{bias dep}(d),  and obtain $v_{\text{s}}$. At $I$=0, there is a very small spin-signal $v_{\text{s}}\sim$ 3 nV (black dash line in Fig.~\ref{bias dep} (d)). On applying $I$ across the injector electrode,  in line with the previous studies on Gr-hBN tunnel barrier systems \cite{gurram_bias_2017, leutenantsmeyer_efficient_2018},  $v_{\text{s}}$ increases in magnitude and changes its sign on reversing the polarity of $I$ (Fig.~\ref{bias dep} (d)). Similarly, we also measure the $2^{\text{nd}}$ and $3^{\text{rd}}$ harmonic spin-signals via SV measurements and its bias dependence, as shown in Figs.~\ref{bias dep}(b,c,e,f). The unambiguous measurement of the higher harmonic spin-signals clearly suggests a presence of nonlinear processes in the spin-signal.


To confirm the spin-dependent origin of the nonlinearity in the spin-signal, we perform Hanle spin-precession measurements on $1^ {\text{st}}, 2^ {\text{nd}}$ and $3^ {\text{rd}}$ harmonic spin-signals. Here, for a fixed in-plane magnetization configuration of the injector-detector electrodes (parallel or anti-parallel), as labeled in Figs.~\ref{bias dep}(a,b,c), a magnetic-field $B_{\perp}$ is applied perpendicular to the plane of the device, as shown in Fig.~\ref{dev geo}(a). The injected in-plane spins diffuse towards the detector and precess around $B_{\perp}$ with the Larmor frequency $\omega_{\text{L}}\propto B_{\perp}$.
The whole dynamics is given by the Bloch equation, $D_{\text{s}} {\bigtriangledown}^2\overrightarrow{\mu_{\text{s}}}-\frac{\overrightarrow{\mu_{\text{s}}}}{\tau_{\text{s}}}+\overrightarrow{\omega_{\text{L}}}\times \overrightarrow{\mu_{\text{s}}}=0 \label{bloch}$, with the spin diffusion constant $D_{\text{s}}$, spin relaxation time $\tau_{\text{s}}$, spin-accumulation $\overrightarrow{\mu_{\text{s}}}=v_{\text{s}}/p_{\text{det}}$ in the transport channel, and the spin diffusion length $\lambda_{\text{s}}$ = $ \sqrt{D_{\text{s}}\tau_{\text{s}}}$. The measured 1$^{\text{st}}$,  2$^{\text{nd}}$ and \ 3$^{\text{rd}}$ harmonic Hanle curves are fitted with the solution to the Bloch equation. From the fitting, we consistently obtain $D_{\text{s}}\sim$0.02 m$^2$s$^{-1}$ and $\tau_{\text{s}}\sim$ 650-700 ps resulting in $\lambda_{\text{s}}\sim$ 4 $\mu$m for the 1$^{\text{st}}$ and higher harmonic measurements in Figs~\ref{bias dep}(g-i). Since the spin transport parameters are the same for all harmonics, we conclude  that the higher harmonic spin-signals do not have its origin in the spin-transport process, and pinpoint the origin of the spin-dependent nonlinearity  to the spin-injection process.

To understand the concept of nonlinearity during spin-injection, we now develop an analytical framework. As the differential spin-injection polarization depends on the input dc bias current $I$,  the expression for $p_{\text{inj}}$ using the Tailor expansion around $I=0$ with a small ac charge current $i$ can be written as:
\begin{equation}
p_{\text{inj}}(i)|_{I=0}=p_0(1+C_1i+C_2i^2+...)
\label{nonlin p}
\end{equation}
 where $p_{\text{inj}}=p_0$ in the absence of nonlinear processes, which are enabled via the nonzero constants $C_1,C_2,...$. Now, using Eq.~\ref{vs}, we obtain:
\begin{equation}
v_{\text{s}}\propto p_0i+p_0C_1i^2+...
 \label{master eq}.
\end{equation}
which enables us to measure the presence of higher harmonic spin-signals $\propto i^2, i^3,...$ due to the nonlinearity introduced by the spin-injection process in Eq.~\ref{nonlin p}.


As shown in the spin-transport measurements in Fig.~\ref{bias dep}, the nonlinearity can be experimentally probed by using the mixed signal (ac+dc) measurements.  When an input current $i+I$ is applied to such nonlinear system, the expression for the $1^{\text{st}}$ harmonic spin-signal $v_{\text{s}}$, obtained by replacing $i$ with $i+I$ in Eq.~\ref{master eq},  acquires a different function form (see Supplementary Material for derivation) and contains a bias $I$ dependent term : 
\begin{equation}
v_{\text{s}}\sim\{p_0(1+2C_1I) \}R_{\text{s}}p_{\text{det}}i.
\label{hmc}
\end{equation} 
As a consequence of the nonlinearity present in the spin-signal  (Eq.~\ref{master eq}), additional terms with the mixing of $i$ and $I$ appear, and now $p_{\text{inj}}\sim p_0(1+2C_1I)$ is obtained instead of $p_0$ (at  $I=0$). For such case, one would expect a gain in $p_{\text{inj}}\propto I$. Indeed, corroborating with the hypothesis in Eq.~\ref{hmc}, $v_{\text{s}}$ increases in magnitude with the applied dc bias $I$ and reverses its sign with the dc current polarity (Fig.~\ref{bias dep}(d)).

Similarly, the expressions for $n^{\text{th}}(n\geq 2)$ harmonic  components of $v_{\text{s}}\propto (C_{n-1}+C_nI)i^n$ due to nonzero $C_j$s are obtained using the mixed signal analysis (see Supplementary Material for detailed expressions). Here, the contribution from $C_{n-1}$, i.e. the $n^{\text{th}}$ order term in Eq.~\ref{master eq} would appear even if only the ac current is applied.   In presence of a nonzero $I$, the higher order term $C_{n}$  would also contribute to the $n^{\text{th}}$ order spin-signal and introduce the dc bias dependence on the spin-signal.  

For SV measurements in our device, we can only measure the even harmonic spin-signal , i.e. $2^{\text{nd}}$ (Fig.~\ref{bias dep}(b,e)) and $4^{\text{th}}$ (Supplementary Material) harmonic using the pure ac current injection ($I=0$).  However, similar to the $1^{\text{st}}$ harmonic spin-signal, higher odd ($3^{\text{rd}}$) harmonic spin-signal (Fig.~\ref{bias dep}(c,f)) can be measured unambiguously only with the application of the dc bias.  When a nonzero $I$ is applied, the contribution of even harmonic signals couples to the odd harmonic spin-signals, and now the odd harmonic responses can also be measured. The dominance of only even harmonic components in the spin-signal is peculiar, and is not clear at the moment. Also, the bias-dependent behaviour of higher harmonic spin-signals can be explained via the expressions obtained from the mixed-signal analysis only near the zero-bias, where higher ($\geq$ 5$^{\text{th}}$) harmonic components do not play a major role. A complete understanding of this behaviour warrants the inclusion of higher order terms in the expression for contact polarization as well as higher harmonic SV measurements for the estimation of the proportionality constant $C_j$($\geq$5$^{\text{th}}$ harmonic).

\subsection{Analog signal-processing of spin-signal due to nonlinear effects}

\begin{figure*}
\includegraphics[scale=1]{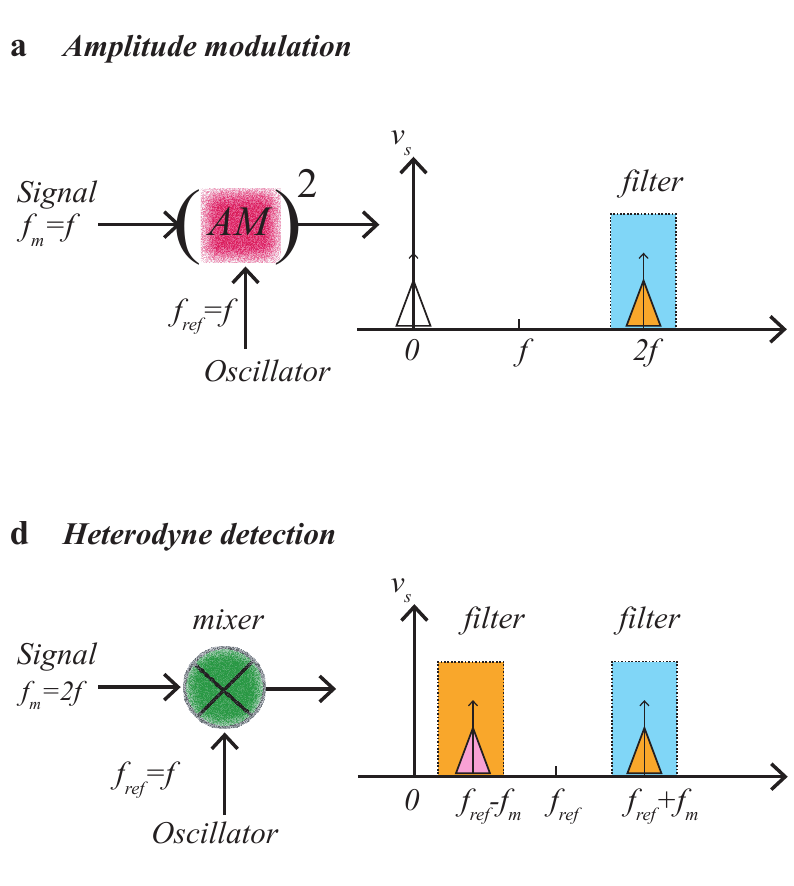}
\includegraphics[scale=1]{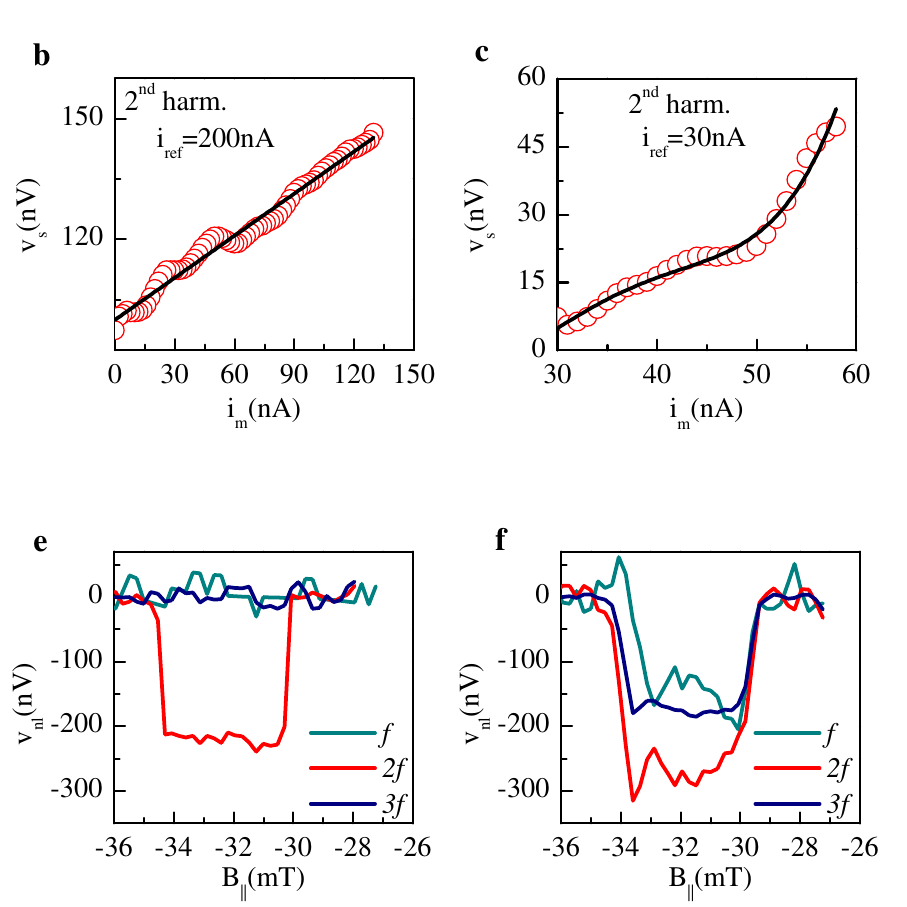} 
\caption{\label{am fm} \textbf{Analog spin-signal processing} \textbf{a.} Amplitude modulation (AM) scheme. \textbf{b.} AM measurement of the spin-signal $v_{\text{s}}$  for $i_{\text{m}}<< i_{\text{ref}}$ and the linear fit (black) \textbf{c.} $i_{\text{m}}\geq i_{\text{ref}}$ and the nonlinear fit (4th order polynomial).  \textbf{d.} Heterodyne detection scheme \textbf{e.} for $i_{\text{m}}=0$  no spin-signal is present at the frequency $f$ and $3f$. \textbf{f.} For $i_{\text{m}}\neq0$, due to the frequency-shifting of the signal present at $2f$ to $f$ and $3f$, equal strength signal appears at both frequencies. Both measurements were performed at RT. SV measurements in (e) and (f) are offset to zero for clear representation.}
\end{figure*}
The presence of nonlinearity which gives rise to signal-amplification , is fundamental to analog signal-processing operations \cite{razavi}. In our spintronic device, we exploit the spin-dependent nolinearity  and demonstrate its applications straightaway by performing the spin analogues of well established analog electronic operations. 

\subsubsection{Amplitude modulation}
 
For amplitude modulation (AM) signal-processing \cite{oppn}, a modulating input  $i_{\text{m}}$ along with a reference input  $i_{\text{ref}}$, both having the same frequency $f$, are applied to a nonlinear element (Fig.~\ref{am fm}(a)). As a result of signal mixing, for our nonlinear spintronic device, the output spin-signal $v_{\text{s}} \propto (i_{\text{ref}}+i_{\text{m}})^2$ is detected at frequency $2f$.  For a constant $i_{\text{ref}}$, if $i_{\text{m}}<<i_{\text{ref}}$,  the measured spin-signal will be $\propto i_{\text{ref}}i_{\text{m}}$, implying the effect will be linear in $i_{\text{m}}$ at the detection frequency $2f$( $2^{\text{nd}}$ harmonic response), and we can realize an analog spin-signal multiplier. 

In order to measure this effect, we inject $i_{\text{ref}}$ = 200 nA and modulate $i_{\text{m}}$ in the range of 0-120 nA (both at $f$=7 Hz) at the injector (Fig.~\ref{dev geo}(a)). We measure the $2^{\text{nd}}$ harmonic $v_{\text{s}}$ via SV measurements. The measured spin-signal  is linear in $i_{\text{m}}$ (Fig.~\ref{am fm}(b)) and thus the device acts as a spin-signal multiplier. For the other situation, i.e. when $ i_{\text{m}}>>i_{\text{ref}}$,  $v_{\text{s}}\propto (i_{\text{m}})^2$. In this case, we fix $i_{\text{ref}}$=30 nA and modulate $i_{\text{m}}$ in the range 30-60 nA. The measured response of $v_{\text{s}}$ (Fig.~\ref{am fm}(c)) clearly deviates from the earlier measured linear response in Fig.~\ref{am fm}(c). However, due to the contribution of higher-order terms to the $2^{\text{nd}}$ harmonic signal, the measurement in Fig.~\ref{am fm}(c) is better explained by the $4^{\text{th}}$ order polynomial fit instead of a parabolic fit.

\subsubsection{Heterodyne detection} As another demonstration of signal-processing, in a heterodyne detection method the input signal frequencies are not equal, i.e. $f_{\text{m}}\neq f_{\text{ref}}$, and one obtains the signal at the heterodyne frequencies $f_{\text{ref}}\pm f_{\text{m}}$  at the output of the nonlinear element \cite{sedra_smith,oppn}.  In order to realize this operation, $i_{\text{ref}}$ at the frequency $f_{\text{ref}}=f$ and $i_{\text{m}}$ at $f_{\text{m}}=2f$ are applied at the injector input (Fig.~\ref{am fm}(d)). The nonlinear component of the spin-signal $v_{\text{s}}$ is $\propto (i_{\text{ref}}\sin(2\pi ft)+i_{\text{m}}\sin(2\pi 2ft))^2$. If $i_{\text{m}}$ =0, one would expect the spin-signal $v_{\text{s}}$ at $2f$. Interestingly, for $i_{\text{m}}\neq 0$, $v_{\text{s}}\propto i_{\text{ref}}i_{\text{m}}$ can also be detected at the $1^{\text{st}}(f=2f-f)$ and $3^{\text{rd}}(3f=2f+f)$ harmonic components.

In our measurements, for $i_{\text{m}}$=0 and $i_{\text{ref}}$=200 nA ($f$=7 Hz), only the $2^{\text{nd}}$ harmonic spin-signal is measured ( Fig.~\ref{am fm}(e)). When we also apply $i_{\text{m}}$=150 nA  at the input frequency $2f$, spin-valve signals of similar magnitudes are detected at frequencies both at $f$ and $3f$ ( Fig.~\ref{am fm}(f)), which is a clear demonstration of $heterodyne$ detection of spin-signals. Note that earlier there was no measurable odd ($1^{\text{st}}$ and $3^{\text{rd}}$) harmonic  spin-signal at $I=0$ (Figs.~\ref{bias dep}(d,f)) due to low injection-polarization/ high-noise present in the signal. Now, using the heterodyne detection method we can clearly measure $v_{\text{s}}$ in the $1^{\text{st}}$ harmonic even without applying $I$. In fact, this effect is equivalent to applying a $dc$ current, as both heterodyne and ac+dc measurements couple the higher harmonic spin-signals to the $1^{\text{st}}$ harmonic spin-signal. This method can be used to detect spin-signals at low frequencies where the spin-dependent noise would dominate in spintronic circuits \cite{omar_spin_2017,omar_two-channel_2017}.  Furthermore, the method can also be used as an electrical analog of the heterodyne detection in the field on optical spin-noise spectroscopy \cite{cronenberger_quantum_2016, sterin_optical_2018}.

\subsection{Nonlinear spin-to-charge conversion}
\begin{figure*}
\includegraphics[scale=1]{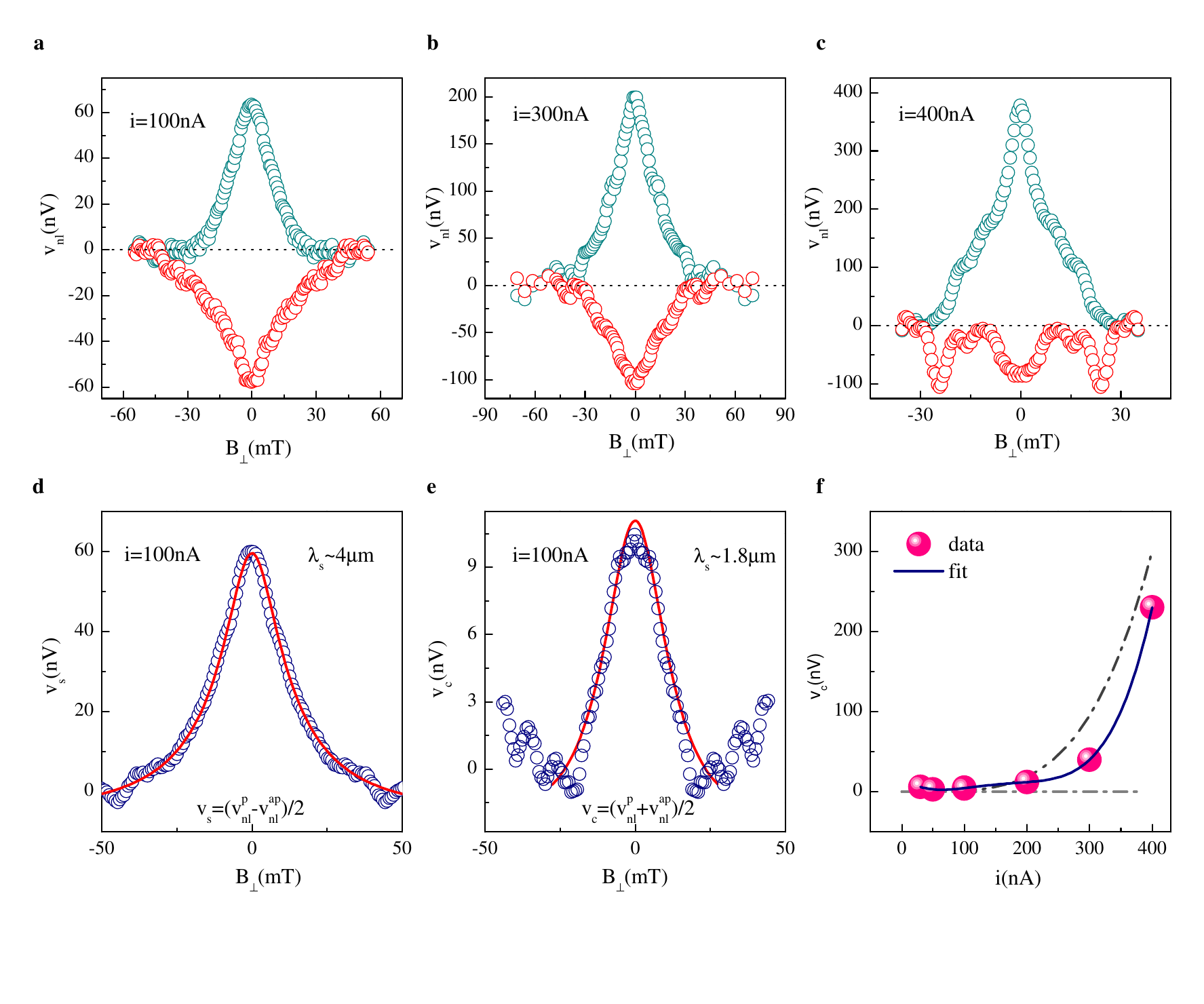}
\caption{\label{nl s2c} \textbf{Nonlocal $2^{\text{nd}}$ harmonic spin-to-charge conversion.}  $2^{\text{nd}}$ harmonic Hanle measurements in the parallel (green) and anti-parallel (red) configuration for $i_{ac}$=\textbf{a.} 100 nA, \textbf{b.} 300 nA and \textbf{c.} 400 nA. The enhancement in the spin-accumulation induced charge signal $v_{\text{c}}$   with $i$ is visible in the asymmetry between $v_{\text{nl}}^{\text{p}}$ and $v_{\text{nl}}^{\text{ap}}$ with respect to the spin-independent background (dashed black line).   The measured data is symmetrized in order to remove the spin-independent linear component present in the data and is offset to zero. All measurements are performed at 4K.    $2^{\text{nd}}$ harmonic Hanle spin-precession measurements  of \textbf{d.} the spin-signal $v_{\text{s}}$ and \textbf{e.}  the spin-accumulation induced charge-signal $v_{\text{c}}$. Both data are obtained from (a) for $i$=100 nA and fitted with the solution to the Bloch equation.  From the fit in (d) and (e), we obtain $\lambda_{\text{s}}\sim$ 4 $\mu$m and 1.8 $\mu$m for the spin and charge signal, respectively. \textbf{f.} $v_{\text{c}}-i$ dependence for the $2^{\text{nd}}$ harmonic data is fitted (blue curve) with a $4^{\text{th}}$  order polynomial function. The dark(light)-gray dashed line is the calculated magnitude of $2^{\text{nd}}$ harmonic component of $v_{\text{c}}$ due to the nonlinear(linear) spin-injection. }
\end{figure*}

So far we have demonstrated that the nonlinearity present in the spin-signal in a Gr/hBN heterostructure has its origin in the spin-injection process, not in the spin transport parameters. However, the nonlinearity in the spin-injection has an important consequence, and can amplify another nonlinear effect present in a small magnitude, i.e. spin-to-charge conversion \cite{vera-marun_nonlinear_2011, vera-marun_nonlinear_2012} in the spin-transport channel. The effect requires the energy dependent conductivity of the transport channel and the presence of spin-accumulation as prerequisites. A nonlocal charge-signal $v_{\text{c}}$ due to energy-dependent spin-to-charge conversion is given by:

\begin{equation}
v_{\text{c}}= C_0\mu_{\text{s}}^2=C_0{(p_{\text{inj}}R_{\text{s}}e)^2}i^2,
\label{vc}
\end{equation}
where $C_0$ is a proportinality constant (see Supplementary Material for details), and we have used the relation $\mu_{\text{s}}=v_{\text{s}}e/p_{\text{det}}$  and Eq.~\ref{vs} to obtain the $v_{\text{c}}$-$i$ dependence.

Now, to probe the spin-to-charge conversion effect and the spin-dependent origin of the nonlocal charge voltage, we perform Hanle measurements. Since the spin-to-charge conversion is a $2^{\text{nd}}$ harmonic effect for the applied charge current $i$, we inject a pure ac current $i$ in the range 100-400 nA and measure the $2^{\text{nd}}$ harmonic response of  $v_{\text{nl}}^{\text{p(ap)}}$ as a function of $B_{\perp}$ using the measurement geometry in Fig.~\ref{dev geo}(a).   In our measurements, we observe an asymmetry between the magnitudes of $v_{\text{nl}}^{\text{p}}$ and $v_{\text{nl}}^{\text{ap}}$ in Figs.~\ref{nl s2c}(a-c), which is present in a small magnitude for $i$=100 nA and grows rapidly for $i$=400 nA to such extent that the Hanle-dephasing of $v_{\text{nl}}$ is measured properly only in the parallel configuration.

 To understand the origin of this asymmetry, we plot the nonlocal charge voltage $v_{\text{c}}=(v_{\text{nl}}^{\text{p}}+ v_{\text{nl}}^{\text{ap}})/2$ ( Fig.~\ref{nl s2c}(e)) and the spin-signal $v_{\text{s}}=(v_{\text{nl}}^{\text{p}}- v_{\text{nl}}^{\text{ap}})/2$ ( Fig.~\ref{nl s2c}(d)). The Hanle like shape of $v_{\text{c}}(B_{\perp})$  in Fig.~\ref{nl s2c}(e) immediately confirms that indeed the nonlocally measured charge voltage $v_{c}$  has the spin-dependent origin, and is reduced to zero in the absence of spin-accumulation (at $B_{\perp}\sim$ 40 mT).  Next, due to its square dependence on $\mu_{s}$ in Eq.~\ref{vc}, $v_{\text{c}}$ should decay with the characteristic spin-relaxation length $\lambda_{\text{s}}/2$ instead of $\lambda_{\text{s}}$ \cite{vera-marun_nonlinear_2011}. In order to verify this hypothesis (Eq.~\ref{vc}), we fit $v_{\text{c}}$-$B_{\perp}$ dependence in  Fig.~\ref{nl s2c}(e) with the solution to the Bloch equation, and obtain $\lambda_{\text{s}}\sim$ 2 $\mu$m, which is about half the spin relaxation length obtained via the Hanle spin-precession measurements on the spin-signal $v_{\text{s}}$ (Fig.~\ref{nl s2c}(d)). The same effect appears in the $1^{\text{st}}$ harmonic $v_{\text{c}}$  due to its coupling with the $2^{\text{nd}}$ harmonic effect in presence of a nonzero $I$ (see Supplementary Material). In this way, we unambiguously establish the spin-dependent origin and the square dependence of the nonlocal charge voltage on spin-accumulation via Hanle measurements.

 Lastly, the $v_{\text{c}}$-$i$ dependence  is plotted in Fig.~\ref{nl s2c}(f).  The dark (light) grey dashed line is the calculated magnitude of $v_{\text{c}}$ while considering the contribution from nonlinear (linear) spin injection with  $i^4(i^2)$ dependence on the injected current  (see Supplementary Material for details). The measured data is in close agreement with the calculated $v_{\text{c}}$ due to the nonlinear spin-injection, and is better fitted with a $4^{\text{th}}$ order polynomial than a parabolic function. Clearly, such efficient spin-to-charge conversion cannot be explained only via the linear spin-injection process, and the contribution from the nonlinear processes has to be taken into account. In conclusion, Gr/hBN heterostructures  due to  the presence of nonlinear spin-injection offer a highly efficient platform to probe nonlinear spin-to-charge conversion effect. The interaction of the two nonlinear effects  produces a measurable effect without needing any additional effect such as  spin-orbit coupling \cite{safeer_room-temperature_2019, ghiasi_charge--spin_2019}.

To summarize, we, for the first time demonstrate the presence of spin-dependent nonlinearity in a spintronic device via all electrical measurements. This effect is the key ingredient in signal-processing, and opens up the portal for the development of the field of analog spintronics, following the pathway of the electronic revolution.  Our results suggest that nonlinearity can be exploited in multiple ways to manipulate spin-information such as via complex signal-processing and spin-to-charge conversion, and  develop advanced multi-functional spintronic devices  \cite{behin-aein_proposal_2010, vera-marun_nonlinear_2012,vera-marun_nonlinear_2011, acremann_amplifier_2008,  gurram_electrical_2018} and spin-based neuromorphic computing \cite{torrejon_neuromorphic_2017}.

\section{Methods}
\subsection{a. Sample fabrication} We prepare a fully hBN encapsulated graphene stack via a dry pick-up transfer method. The hBN  (thickness$\sim$ 7 nm) and graphene flakes are exfoliated on SiO$_{\text{2}}$/Si substrate and identified via optical contrast analysis using an optical microscope. The thickness of hBN layer is measured via the atomic force microscopy and is $\sim$ 9nm for the bottom hBN substrate and 0.9-1.0 nm (3L) for the top hBN layer. For the stack preparation, the 3L-hBN flake is brought in contact with a visco-elastic PDMS (polydimethylsiloxane) stamp which has a sticky PC (polycarbonate) film attached to it. During the contact, the whole assembly is heated and the hBN is picked up by the PC film. Following the same step, the Gr flake is picked up by the hBN flake on the PC film due to the van der Waals interaction between these two layers. In the last step, the thick hBN flake on the SiO$_{\text{2}}$ substrate is brought in contact with  
 the Gr/3L-hBN on the PC film, the whole assembly is heated up to 150$^{\circ}$C and the PC film with the Gr/3L-hBN is released on to the bottom hBN substrate. Afterward, the bottom-hBN/Gr/3L-hBN stack is put in chloroform solution at room temperature to dissolve the PC film. In order to remove the remaining polymer residues on top of the top 3L-hBN layer, the stack is annealed at 250$^{\circ}$C in Ar-H$_{\text{2}}$ environment for 7 hours.

 \subsection{b. Device Fabrication}
  The electrodes are patterned via the electron-beam lithography on the PMMA (poly-methyl methacrylate) spincoated  sample. Then the sample is developed in a MIBK:IPA solution for 60 seconds in order to remove the polymer from the electron-beam exposed area. Next, to obtain the spin-sensitive electrodes, 65 nm thick cobalt is deposited on the sample via electron-beam evaporation. In order to prevent the oxidation of cobalt, a 3 nm thick layer of aluminium is deposited on top. The residual metal on top of the polymer is removed by performing the lift-off in hot acetone at 40$^{\circ}$C. 
 
\subsection{c. Measurements}

Measurements were performed both at 4K (Helium temperature) and room temperature in vacuum in a flow cryostat. Differential ac signal measurements were performed using low frequency lock-in detection method. For mixed signal (ac+dc) measurements and back-gate application, Keithley 2410 dc source was used.  


 \section{Acknowledgements}
We acknowledge  J. G. Holstein, H.H. de Vries, T. Schouten, H. Adema and A. Joshua for their technical assistance. We thank B.N. Madhushankar, RUG for the help in sample preparation, and  A. Kamra, NTNU for critically reading the manuscript. Growth of hexagonal boron nitride crystals was supported by the
Elemental Strategy Initiative conducted by the MEXT, Japan and the CREST(JPMJCR15F3), JST.  MHDG acknowledges financial support from the Dutch Research Council (NWO VENI 15093). This research work was funded by the the Graphene flagship core 1 and core 2 program  (grant no. 696656 and 785219), Spinoza Prize (for B.J.v.W.) by the Netherlands Organization for Scientific Research (NWO) and supported by the Zernike Institute for Advanced Materials.

\section{Author Contributions}

S.O., M.G. and B.J.v.W. conceived the experiment. S.O., K.W. and T.T. carried out the sample fabrication. S.O. and M.H.D.G. carried out the experiment. S.O., M.G.,M.H.D.G. and B.J.v.W. carried out the analysis and wrote the manuscript. All authors discussed the results and the manuscript.

 

\newpage

 \beginsupplement

\begin{center}
 \textbf{\large Supplementary Information}
\end{center}

\section{I. Model for nonlinear spin-injection}

Current/voltage bias dependence of the spin-injection efficiency across the cobalt/hBN/Gr contact introduces nonlinearity in the spin-injection process. As discussed in the main text, the differential polarization depends on the input bias current $I$,  the expression for $p_{\text{inj}}$ using the Tailor expansion around $I=0$ can be written as:
\begin{equation}
p_{\text{inj}}(i)=p_0(1+C_1i+C_2i^2+...)
\label{bias polarization}
\end{equation}
where $p_0$ is the unbiased differential contact polarization. The spin-signal in the nonlocal geometry is:
\begin{equation}
 v_{\text{s}}=\{p_{\text{inj}} \}\times\ R_{\text{s}}  \times p_{\text{det}}i.
 \label{vnl}
\end{equation}
Here, $p_{\text{det}}$ is the unbiased detector polarization, $R_{\text{s}}$ is the effective graphene spin resistance. By substituting Eq.\ref{bias polarization} into Eq.\ref{vnl}, we obtain the following expression for $v_{\text{s}}$:

\begin{equation}
\begin{split}
 v_{\text{s}}=&\{p_0(1+C_1i+C_2i^2+...) i\}\times R_{\text{s}} \times p_{\text{det}}\\
  =&\{p_0 \times R_{\text{s}} \times p_{\text{det}}\}i+\{p_0C_1 \times R_{\text{s}} \times p_{\text{det}}\}i^2\\
&+\{p_0C_2 \times R_{\text{s}} \times p_{\text{det}}\}i^3+...
  \end{split}
 \label{master eq supp}
\end{equation}
 Due to the presence of  nonlinearity in the spin-injection process, even the 2$^{\text{nd}}$, 3$^{\text{rd}}$ and 4$^{\text{th}}$ harmonic spin-signals can be measured. 

\section{II. Mixed-signal analysis}
The nonlinearity in the spin-signal can  be measured via mixed-signal analysis which is a well established framework in the field of analog electronic-circuit design. We use this tool to measure and analyze the nonlinearity present in the spin-injection process. 

When two independent inputs are supplied to a linear system, its response will be a linear combination of the input signals. However, this is not the case when the system possesses nonlinearity, and additional contributions would appear due to the mixing of the independent input signals. 
In order to measure the response of the ac and dc input signals and their mixing, we apply a charge current $I+i\sin \omega t$. We apply a dc current using a home built dc current source and the ac current using a lock-in source  across the injector electrodes. Since we know from the measurements presented in the main text that $R_{\text{s}}$ and $p_{\text{det}}$ remain constant, for sake of simplicity we omit the constants in Eq.\ref{master eq supp}, and to explain the analysis, we assume $p_{\text{inj}}=p_0(1+C_1I)$ and omit the higher order terms. Now, $v_{\text{s}}$ is: 
\begin{equation}
\begin{split}
 v_{\text{s}}&\backsimeq\{p_0(1+C_1(I+i\sin\omega t))\}\times (I+i\sin\omega t)\\
&\backsimeq\{p_0 (I+i\sin\omega t)\}+p_0C_1\times\{(I+i\sin\omega t)^2\}\\
&\backsimeq\{p_0 (I+C_1I^2)\}+\{p_0(1+2C_1I)i\sin\omega t\}\\
&+\{ p_0C_1i^2\sin^2\omega t\}
 \end{split}
 \label{vnl hmc}
\end{equation}

$v_{\text{s}}$ in Eq.~\ref{vnl hmc} has three distinct contributions, a dc, a $1^{\text{st}}$ harmonic (i.e. $\propto \sin\omega t$) and a $2^{\text{nd}}$ harmonic ($\propto \sin^2\omega t$) contribution, separated in curly brackets. In a lock-in measurement,  if an ac current $i$ is applied to the sample at a lockin reference frequency $f=\frac{\omega}{2\pi}$, only the components of $v_{\text{s}}$ appearing at frequency $f$ or at higher harmonics $2f, 3f, ...$ would  be measured via the lock-in detection method. Other contributions  are filtered out and are not measured via the lock-in amplifier. In order to do so, the smaple output $v_{\text{s}}$ in Eq.~\ref{vnl hmc} at the lock-in input is multiplied with the reference signal $\propto \sin n\omega t$, where $n=1,2,...$  in order to measure the $1^{\text{st}}$, $2^{\text{nd}}, ...$ harmonic contributions. Then the output is low-pass filtered to obtain a dc output. 

In order to filter out the $1^{\text{st}}$  harmonic contribution, $v_{\text{s}}$ in Eq.~\ref{vnl hmc} is multiplied with the reference signal $\sin\omega t$, and $v_{\text{s}}\propto \sin\omega t$ is only filtered out and measured in the $1^{\text{st}}$ harmonic response:
\begin{equation} 
v_{\text{s}}=\{p_0(1+2C_1I)\}\times R_{\text{s}} \times p_{\text{det}}i
\end{equation}
Here, we would like to emphasize that due to the nonlinear term present in contact polarization, i.e. because of a  nonzero $C_1$, contact polarization is not equal to $p_{\text{inj}}=p_0$ anymore and is modified to $p_{\text{inj}}=p_0(1+2C_1I)$. Therefore, in presence of the nonlinearity the dc and differential contact polarization will not be equal and the differential one may exceed the dc polarization \cite{gurram_bias_2017}. 

Now, in the same way, the $2^{\text{nd}}$ harmonic component of $v_{\text{s}}$ can be filtered out by multiplying $v_{\text{s}}$ in Eq.~\ref{vnl hmc} with $\sin2\omega t$:
\begin{equation}
v_{\text{s}}=\{p_0C_1\}\times  R_{\text{s}}  \times p_{\text{det}}i^2
\label{vs 2nd hmc}
\end{equation}

It is evident from the expression in Eq.~\ref{vs 2nd hmc} that if $C_1\neq0$, the spin-valve effect would be observed in the $2^{\text{nd}}$ harmonic measurements as well.

 Using the analysis presented above, higher order nonlinearity can be included in the contact polarization $p_{\text{inj}}$ with nonzero $C_2,...$ as in Eq.~\ref{bias polarization} in the same way. Without the loss of generality, following the arguments presented above, one would expect the spin-valve effect to appear in the $3^{\text{rd}}$ harmonic measurements for $C_2\neq 0$ and the $C_2I$ dependent terms in the $1^{\text{st}}$ and $2^{\text{nd}}$ harmonic spin-signals, as can be seen in  Fig.2 of the main text. In this way, for highly nonlinear spin-injection processes, there will be contributions of higher order spin-injection processes appearing in the low-order terms due to the coupling of higher order processes with the charge current. 

\section{III. nonlinear spin-to-charge conversion}
Ferromagnets (FM) have a nonzero spin-dependent conductivity $\sigma_{\text{s}}$, i.e. spin-up and spin-down electrons in FM materials have different conductivity $\sigma^{\uparrow}$ and $\sigma^{\downarrow}$ where $\sigma_{\text{s}}=\sigma^{\uparrow}-\sigma^{\downarrow}$. In presence of a nonzero spin-current, i.e gradient of the spin-accumulation $\mu_{\text{s}}=(\mu^{\uparrow}-\mu^{\downarrow})/2$,  it gives rise to a charge voltage $v_{\text{c}}$ $\propto \sigma_{\text{s}}\nabla \mu_{\text{s}}$  in the FM. However, nonmagnets (NM) have $\sigma_{\text{s}}=0$.

Because of energy-dependent density of states in graphene,  in  presence of a large spin-accumulation $\mu_{\text{s}}$, spin-up and spin-down electrons experience different conductivity  \cite{vera-marun_nonlinear_2011, vera-marun_nonlinear_2012}. Therefore, in spite of being a nonmagnet, graphene develops a nonzero spin-polarization $P_{\text{d}}$ away from the Fermi level, and behaves as a $pseudo$-ferromagnet. As a consequence, a  nonlocal charge voltage $v_{\text{c}}$  is developed along the spin-transport channel length:
\begin{equation}
v_{\text{c}}=-P_{\text{d}}(\mu_{s}/e)
\end{equation}
 where $P_{\text{d}}$ is the spin-to-charge conversion efficiency and can be represented as:
\begin{equation}
P_{\text{d}}=-\mu_{\text{s}} \times \frac{1}{\sigma} \times \frac{\delta \sigma}{\delta E}
\end{equation}
 where $\sigma$ is the energy-dependent conductivity of graphene. Now, $v_{\text{c}}$ can be written as :
\begin{equation}
\begin{split}
v_{\text{c}}&=\frac{1}{\sigma} \times \frac{\delta \sigma}{\delta E} \times \frac{{(\mu_{\text{s}})}^2}{e} \\
&=C_0\mu_s^2\propto i^2
\end{split}
\label{eq3}
\end{equation}

\begin{figure}
\includegraphics[scale=1]{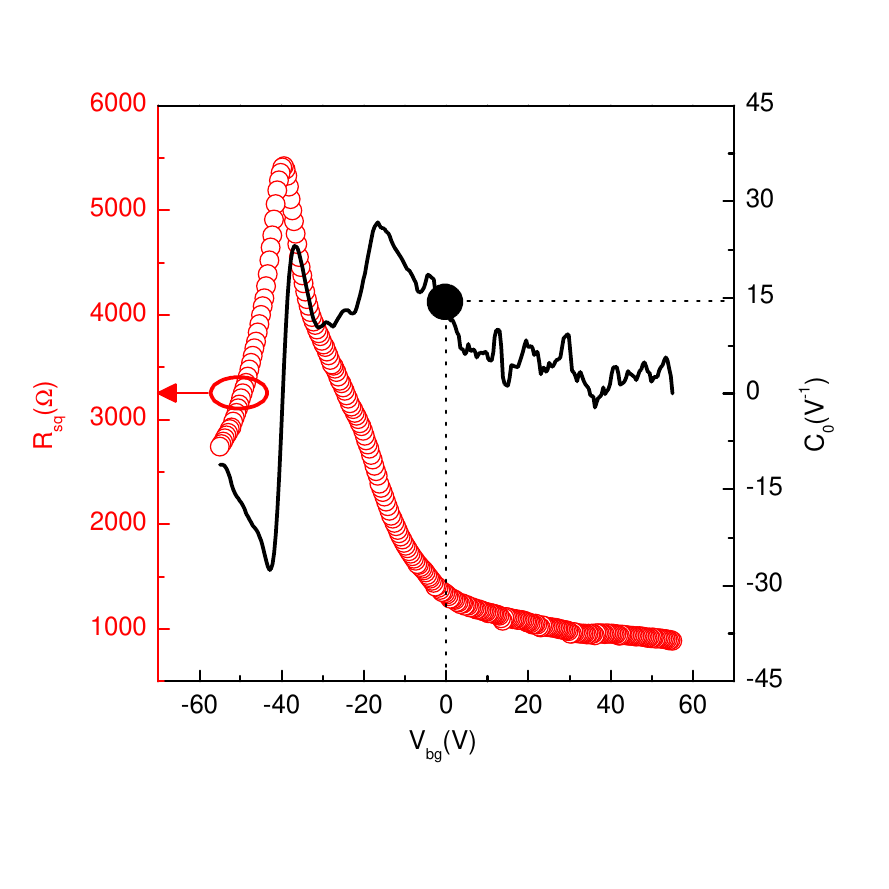}
\caption{\label{c0 value}$R_{\text{sq}}$(left y-axis)-$V_{\text{bg}}$ dependence of graphene (red). On the right y-axis the spin-to-charge conversion efficiency $C_0-V_{\text{bg}}$  dependence (in black) is plotted. $C_0$ at $V_{\text{bg}}$=0V is marked with a solid black dot.  }
\end{figure}

The proportionality constant $C_0$ can be derived from the carrier-density dependent conductivity measurements of graphene and the density of states in graphene (bilayer graphene in our case). The procedure is as follows:

The total number of carriers $n$ can be calculated using the relation:
\begin{equation}
 n=\int_{0}^{E_F}\nu(E)dE
\label{carrier} 
\end{equation}

where the density of states $\nu(E)$ of the BLG is:

\begin{equation}
\nu(E)=\frac{g_sg_v}{4\pi\hbar^2{v_F}^2}(2E+\gamma_1)
\label{dos_blg}
\end{equation}
Here $g_s$ and $g_v$ are electron spin and valley degeneracy(=2), $\hbar$ is the reduced Planck coefficient, $v_F$=10$^6$ m/s is the electron Fermi velocity, and $\gamma_1$=0.37 eV is the interlayer coupling coefficient. We extract the carrier density in graphene from the Dirac measurements, and use it calculate $\frac{\delta\sigma}{\delta E}$ as a function of $n$ using Eq.~\ref{carrier} and Eq.~\ref{dos_blg}. Then, we can easily estimate $C_0$ using Eq.~\ref{eq3} for our sample. Using this procedure, we obtain $C_0$ as a function of the back-gate voltage $V_{\text{bg}}$ (Fig.~\ref{c0 value}). Since, we perform all measurements at $V_{\text{bg}}$=0 V, we use $C_0\sim$15 V$^{-1}$ for further calculations.

\section{III. Model for spin to charge conversion}

We know, from Eq.~\ref{eq3}, the spin-accumulation induced charge voltage $v_{\text{c}}=C_0\mu_{\text{s}}^2$. If an ac charge current $i$ is applied across the injector electrode, it creates a spin accumulation $\mu_{\text{s}}=p_{\text{inj}}iR_{\text{s}}$ underneath the injector electrode, which is measured at a distance $L$ away from the injector electrode. 

\begin{equation}
\begin{split}
v_{\text{c}}&=C_0\mu_{\text{s}}^2=C_{0}(p_{\text{inj}}iR_{\text{s}}e)^2\\
&=C(p_{\text{inj}}i)^2
\end{split}
\label{vc nl1}
\end{equation}
where $C=C_0(R_{\text{s}}e)^2$.  While deriving the expression for $v_{\text{c}}$, we also consider the role of the nonlinear spin-injection and assume $p_{\text{inj}}=p_0(1+C_1i)$, and substitute this expression into Eq.~\ref{vc nl1}:
\begin{equation}
\begin{split}
v_{\text{c}}&=Cp_0^2(1+C_1i)^2i^2\\
&=Cp_0^2(i^2+2C_1i^3+C_1^2i^4)\\
\end{split}
\label{vc nl2}
\end{equation}
Now, we can perform the mixed signal analysis on the expression in Eq.~\ref{vc nl2} by replacing $i$ with $I+i\sin \omega t$ in the same way as described in supplementary section II and obtain expressions for the  $2^{\text{nd}}$  harmonic components of $v_{\text{c}}$:

\begin{equation}
\begin{split}
v_{\text{c}}^{2^{\text{nd}}}&=Cp_0^2\{i^2+6C_1Ii^2+6C_1^2I^2i^2+C_1^2i^4\}\cos2\omega t\\
\end{split}
\label{vc nl2}
\end{equation}

Here, we would like to remark that for $I=0$, we would expect $v_{\text{c}}^{2^{\text{nd}}} \propto i^2$ for small $i$ and $v_{\text{c}}^{2^{\text{nd}}}\propto i^4$ for the large $i$ values. The consequences of this dependence are significant as because of a nonzero $C_1$, at large $i$ values, $v_{\text{c}}\propto i^4$ while $v_{\text{s}}\propto i^2$ (Fig.~4 in the main text) . Therefore, at large $i$ the spin-accumulation induced charge-voltage $v_{\text{c}}$ would be comparable to, and can even surpass the spin-signal $v_{\text{s}}$.   

\section{Spin to charge conversion in 1$^{\text{st}}$ harmonic response}


\begin{figure*}
\includegraphics[scale=1]{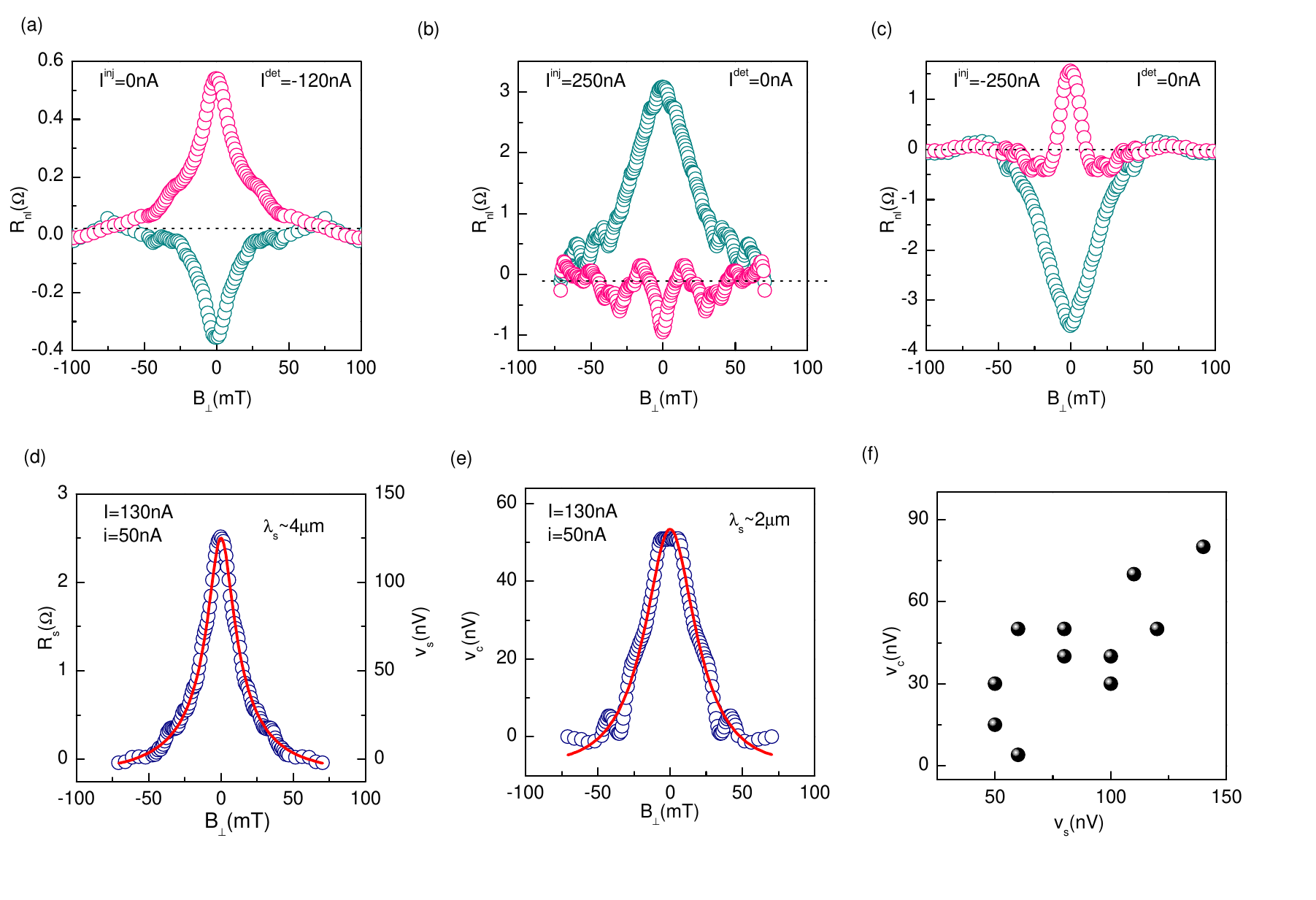}
\caption{ \label{1st hmc s2c}$1^{\text{st}}$ harmonic measurements (a)  $R_{\text{nl}}^{\text{p(ap)}}=\frac{v_{\text{nl}}^{\text{p(ap)}}}{i}$ are roughly symmetric with respect to the background (black dashed line) for $i$=20 $\mu$A and $I=0$ at the injector. Here, the transparent contact C2 is used as an injector and the tunnel contact C1 is used as a detector. When the tunnel contact is used as an injector, a strong asymmetry is measured  between $R_{\text{nl}}^{\text{p}}$ (green) and $R_{\text{nl}}^{\text{ap}}$ (pink) with respect to the background for (b) $I$ = +250 nA and (c) $I$ = -250 nA due to the coupling of the $2^{\text{nd}}$ harmonic spin-to-charge conversion effect to the applied dc bias $I$. $R_{\text{nl}}^{\text{p}}$ (green) is negative in (a) and (c) because of the negative bias current $I$.  $1^{\text{st}}$ harmonic (d) spin-signal $v_{\text{s}}$. (e)  Nonlocal charge-signal $v_{\text{c}}$-$B_{\perp}$ dependence. The data is symmetrized and a constant background is subtracted from the raw data. (f) Summary of $v_{\text{c}}$-$v_{\text{s}}$ dependence. The measurements are performed at 4K.}
\end{figure*}

As explained in Sec.III, spin-to-charge conversion is a nonlinear effect, and is measured as higher harmonic of the nonlocal charge signal. However, similar to the nonlinear spin transport measurements, when a dc  charge current $I$ is injected along with $i$, due to the mixing between $i$ and $I$,  we can also measure  $v_{\text{c}}$  in the $1^{\text{st}}$ harmonic response. The expression for  $1^{\text{st}}$ harmonic component of $v_{\text{c}}$ , obtained by using the mixed signal analysis is:
\begin{equation}
\begin{split}
v_{\text{c}}^{1^{\text{st}}}&=Cp_0^2\{2Ii+6C_1I^2i+(3/2)C_1i^3\\
&+3C_1^2Ii^3+4C_1^2I^3i\}\sin\omega t\\
&\simeq Cp_0^2Ii\{1+3C_1I+2(C_1I)^2\}\hspace{10mm} (i<<I).
\end{split}
\label{vc nl1}
\end{equation}

Here, the terms with $C_1$ in the expression in Eq.~\ref{vc nl1} appear due to the nonlinear spin-injection process. If $C_1I>1$, the term $\propto (C_1I)^2$  will dominate and set amplification factor for the spin-to-charge conversion effect in the $1^{\text{st}}$ harmonic response, which is absent for linear spin-injection. Now, $v_{\text{c}}\simeq C p_0^2 Ii(C_1I)^2$, and due to the nonlinearity present in the spin-injection process, we would expect an amplification in the spin-to-charge conversion effect in the $1^{\text{st}}$ harmonic signal, proportional to $(C_1I)^2$. 

 In order to verify this hypothesis, we revisit the $1^{\text{st}}$ harmonic Hanle measurements, and obtain $v_{\text{c}}$ as an average of $v_{\text{nl}}^{\text{p}}$ and $v_{\text{nl}}^{\text{ap}}$.  We first present the case when no bias is applied at the injector. For I=0, we expect $v_{\text{c}}=0$ and symmetric parallel and anti-parallel Hanle curves. Since we cannot measure any spin-signal clearly while using the tunnel contact $C1$ as an injector without biasing it, due to the presence of large noise and a small spin-signal due to small contact-polarization, we use the transparent contact C2 as a injector and C1 as a detector (Fig.~\ref{dev geo}(a)). Now, C1 can be biased with $I$ to enhance its spin-detection efficiency $p_{\text{det}}$ and the spin-signal can be  measured.  Since $I$=0 at the injector, there should be no coupling between the $1^{\text{st}}$ and $2^{\text{nd}}$ harmonic $v_{\text{c}}$ and the parallel and anti-parallel Hanle signals should be symmetric.  In our measurement while using C2 as an injector and C1 as a detector we measure roughly symmetric Hanle curves with respect to the background (black dashed line) for the parallel and anti-parallel configurations (Fig.~\ref{nl s2c}(d)). However, there is no asymmetry expetected for this case. The reason for such behavior is not clear to us at the moment. 

\begin{figure*}
\includegraphics[scale=1]{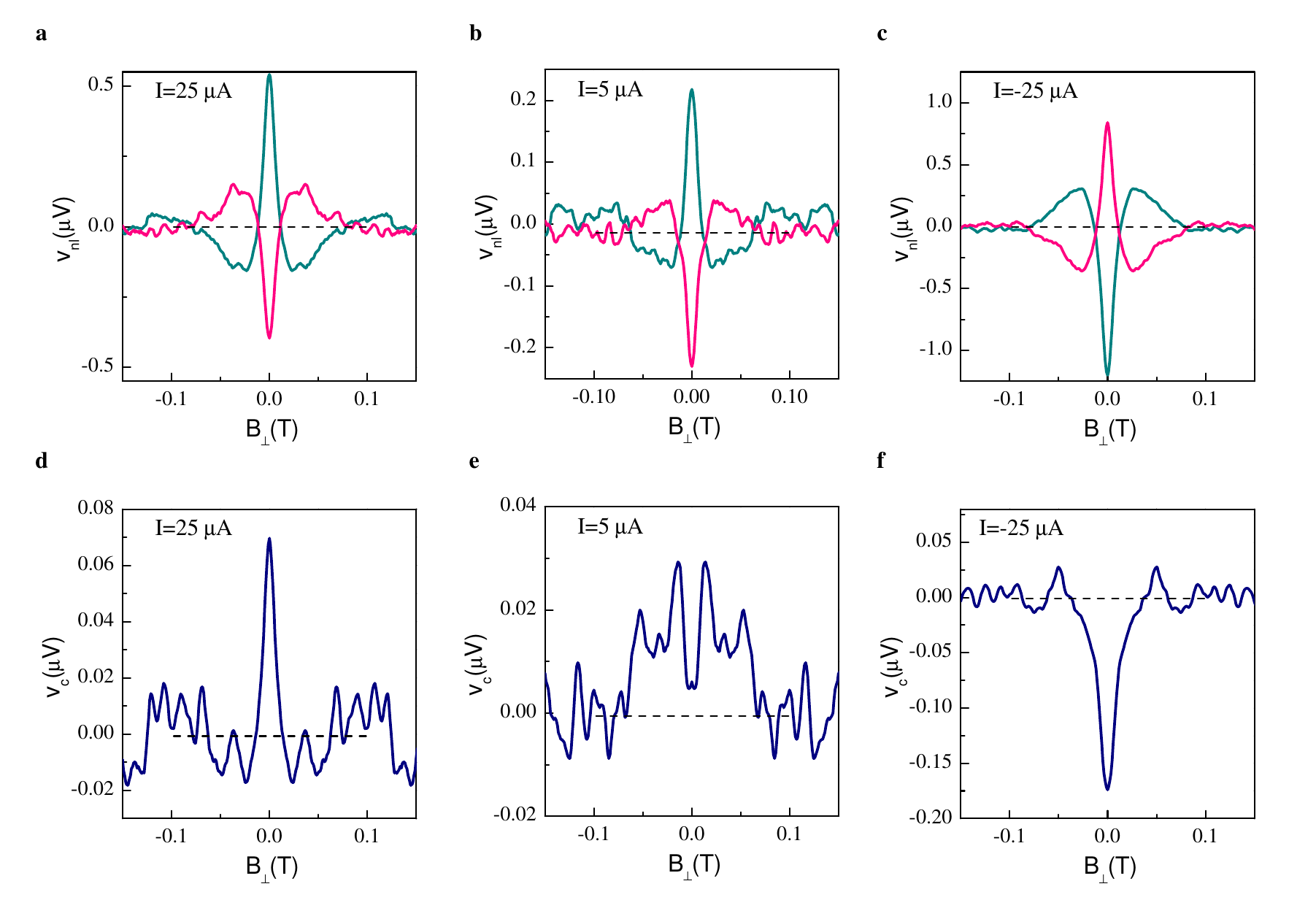}
\caption{ \label{s2c mallik}$1^{\text{st}}$ harmonic measurements for parallel (green) and anti-parallel (pink) configurations, using a bilayer hBN tunnel barrier as a spin-injector at $i$=3 $\mu$A and $I$=(a) 25 $\mu$A (b) 5 $\mu$A  and (c) -25 $\mu$A, and the corresponding $v_{\text{c}}$ at (d), (e) and (f) respectively. The data is symmetrized and a constant background is subtracted from the raw data. The measurements are performed at RT.}
\end{figure*}

Now, when we use the tunnel contact C1 as an injector and C2 as a detector, and inject a charge current $I+i$, as expected we see a huge asymmetry between the magnitudes of parallel and the anti-parallel Hanle signals in Fig.~\ref{1st hmc s2c}(b,c) for $I=\pm250$ nA due to the presence of spin-to-charge conversion effect. Simar to the $2^{\text{nd}}$ harmonic measurements in Fig.~4(c) in the main text, the spin-accumulation induced charge signal which is the average of the parallel and anti-parallel spin-signals  is comparable to the spin-signal. Due to the dominant $2^{\text{nd}}$ order spin-injection and its contribution to the $1^{\text{st}}$ harmonic spin-to-charge conversion, the higher order terms in Eq.~\ref{vc nl1} contribute significantly, and we also see the sign-reversal of $v_{\text{c}}$ with the polarity of $I$. In order to appreciate the role of the nonlinear spin-injection to measure such effect, we also present a similar case for spin-injection  in a different sample  using a 2L-hBN,  in ref.~\cite{gurram_bias_2017} where, where the nonlinear constant $C_1\sim 10^5 A^{-1}$ is not dominant enough compared to $C1 \sim 10^8  A^{-1}$ for 3L-hBN in our sample. Here, even for high enough $I=\pm 25 \mu A$, $v_{\text{c}}$ is significantly small compared to the spin-signal $v_{\text{s}}$ (Fig.~\ref{s2c mallik}), where $v_{\text{c}}$ and $v_{\text{s}}$ are of similar order magnitude.  Therefore, we also do not measure a strong modulation in Hanle shapes in Fig.~\ref{s2c mallik} as in Fig.~\ref{1st hmc s2c} for nonzero $I$. However, the asymmetry between the magnitudes of parallel and anti-parallel Hanle curves, along with the sign reversal in $v_{\text{c}}$ with the polarity of $I$ is consistently present in both measurements. 

 We would also like to remark that the observed behavior is not due to the contribution of outer injector/detector FM electrodes to the measured spin-signal. For nonlocal SV measurements, we consistently observe only two distinct levels in the spin-valve effect corresponding to parallel and anti-parallel configuration of the injector-detector pair. It confirms the contribution of only one FM injector and one detector in contrast with the recently reported two-probe spin-transport measurements in ref.~\cite{gurram_bias_2017, gurram_electrical_2018, spiesser_hanle_2019} where both FM electrodes act as spin-injector and detector contacts and result in asymmetric Hanle curves for parallel and anti-parallel configurations.

Lastly, similar to the $2^{\text{nd}}$ harmonic spin-to-charge conversion effect, shown in Fig.~4(e) of the main text, we also plot the  $1^{\text{st}}$ harmonic $v_{\text{c}}$-$B_{\perp}$ dependence, and obtain the same information. The Hanle-like magnetic-field dependence of $v_{\text{c}}$ in Fig.~\ref{1st hmc s2c}(e) confirms its spin-accumulation induced origin. The fitting of the Hanle curves in  Fig.~\ref{1st hmc s2c}(d) and  Fig.~\ref{1st hmc s2c}(e) results in $\lambda_{\text{s}}\sim$ 4 $\mu$m and $\sim$ 2 $\mu$m, respectively. By fitting the $v_{\text{c}}-B_{\perp}$ dependence in  Fig.~\ref{1st hmc s2c}(e), according to the expectation $v_{\text{c}}\propto v_{\text{s}}^2$,  $\lambda_{\text{s}}\sim$ 2 $\mu$m is obtained which is half of the spin-relaxation length obtained from the spin-signal $v_{\text{s}}$, and again corroborates the square dependence of the nonlocal charge-signal on spin-accumulation, as obtained in the $2^{\text{nd}}$ harmonic measurements in the main text.

\section{4$^{\text{th}}$ harmonic spin signal}

\begin{figure}
\includegraphics[scale=1]{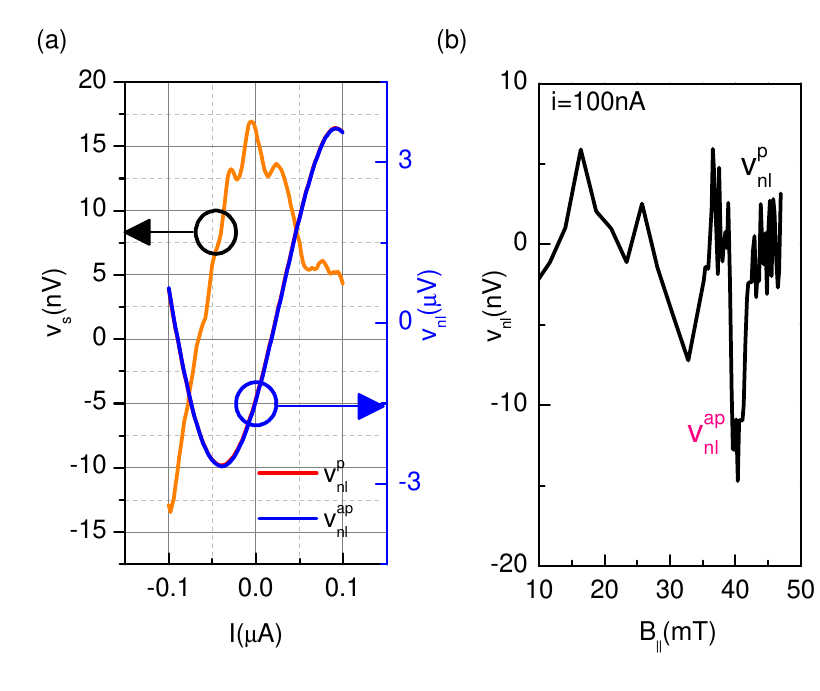}
\caption{\label{4th hmc}(a) 4$^{\text{th}}$ harmonic spin signal (orange) and its bias dependence. The (anti) parallel,  data  in (blue)red is plotted against the right-y axis.  (b) 4$^{\text{th}}$ harmonic spin-valve signal}
\end{figure}

We also measure the $4^ {\text{th}}$ harmonic component of the spin-signal for the input ac current $i$ = 100 nA. The spin-valve effect is shown in Fig.~\ref{4th hmc}(b). Similar to the $2^ {\text{nd}}$ harmonic spin-signal, the $4^ {\text{th}}$ harmonic spin-signal can be measured unambiguously without applying any dc current. As soon as a finite dc bias $I$ is applied along with the ac charge current $i$, again in line with the bias-dependence of the $2^ {\text{nd}}$ harmonic spin-signal (Fig. 2(e) in the main text), the magnitude of the $4^ {\text{th}}$ harmonic signal is also reduced as shown in Fig.~\ref{4th hmc}(a). 




\end{document}